\pdfoutput=1

\documentclass[twocolumn,floatfix,showpacs]{revtex4}
\usepackage{graphicx}
\usepackage{amssymb}
\begin{document}

\title{
  Electron transport and quantum-dot energy levels in
  Z-shaped graphene nanoconstriction with zigzag edges
}

\author{Adam Rycerz}
\affiliation{
  Instytut Fizyki im.\ Mariana Smoluchowskiego, 
  Uniwersytet Jagiello\'{n}ski,
  Reymonta 4, PL--30059 Krak$\acute{o}$w, Poland
}
\affiliation{
  Institut f\"{u}r Theoretische Physik, 
  Universit\"{a}t Regensburg, D--93040 Regensburg, Germany
}

\begin{abstract}
Motivated by recent advances in fabricating graphene nanostructures, we find that an electron can be trapped in $Z$-shaped graphene nanoconstriction with zigzag edges. The central section of the constriction operates as a~single-level quantum dot, as the current flow towards the adjunct sections (rotated by $60^\circ$) is strongly suppressed due to mismatched valley polarization, although each section in isolation shows maximal quantum value of the conductance $G_0=2e^2/h$. We further show, that the trapping mechanism is insensitive to the details of constriction geometry, except from the case when widths of the two neighboring sections are equal. The relation with earlier studies of electron transport through symmetric and asymmetric kinks with zigzag edges is also established. 
\end{abstract}
\date{\today}
\maketitle

\section{Introduction}
Soon after the breakthrough in its fabrication \cite{Gei04,Gei05}, an atomically-thin carbon monolayer (graphene) has attracted intense experimental and theoretical attention. An unusual band structure of graphene leads to exotic electronic properties \cite{Cas09}, which makes possible either to create devices that have no analogue in silicon-based electronics \cite{Gei07}, or to test various predictions of relativistic quantum mechanics in a condensed-matter system \cite{Bee08}. Additionally, graphene's true two-dimensional nature combined with high carrier mobility makes it a promising base material for studying low dimensional systems, such as quantum wires realized as graphene nanoribbons \cite{Che07,Han07,Li08,Wan08,Sta09}, or quantum dots \cite{Sil07,Pon08,Gut08}. Just to give a recent example, the exotic features of quantum chaotic behavior, characteristic for massless spin-$1/2$ fermions confined in a quantum dot \cite{Ber87}, have been found both in an experiment \cite{Pon08} and computer simulations \cite{Wur09a}.

Theoretical research on graphene nanostructures have started much earlier \cite{Fuj96,Nak96,Wak01,Mun06} but speed up after it was realized, that such systems are promising building blocks for a~solid-state quantum computer \cite{Tra07,Ryc07}. In attempt to operate on a~solid-state qubit \cite{Wol01} in graphene, one need to deal with an obstacle that electrons occur in two degenerate families, corresponding to the presence of two different valleys in the band structure. Trauzettel \emph{et al.\/} \cite{Tra07} propose to solve this problem by using the insulating graphene nanoribbon with armchair edges, for which the valley degeneracy is lifted up by the boundary condition \cite{Two06}. Then, applying the gate voltage to a~finite section of the ribbon, one forms the quantum dot which has a~single relevant electronic level (with a spin-only degeneracy) for a considerably wide interval of electron Fermi energies, and thus is called a \textit{single-level quantum dot} (SQD). As a fabrication of perfect armchair nanoribbon seems to be difficult due to the edge instability \cite{Gas08,Hua09} an alternative approaches, employing simultaneously external magnetic field and the mass confinement to break valley degeneracy in graphene rings \cite{Rec07,Zar09}, disks \cite{Rec09}, or antidots \cite{Ped08,She08}, have been discussed. The operational simplicity of the device \cite{Tra07}, offering a~fully-electrostatic control, has inspired designing its counterpart based on a~nanoribbon with predominantly zigzag edges \cite{Wan07}, similar to these already obtained in well-controlled fabrication processes \cite{Han07,Li08,Wan08}. An additional motivation to focus on ribbons with zigzag edges comes from the two theoretical findings: $(i)$ local-density approximation results \cite{Son06} show that the electronic structure narrow graphene constrictions can be well described by a~simple tight binding model, and $(ii)$ analytical discussion of the tight-binding equations for a semi-infinite honeycomb lattice \cite{Akh08a} shows that the zigzag boundary condition applies generically (at low energies) to arbitrary lattice termination, except from the case we have a~perfect armchair edge. 

\begin{figure}[!b]
\centerline{\includegraphics[width=0.7\linewidth]{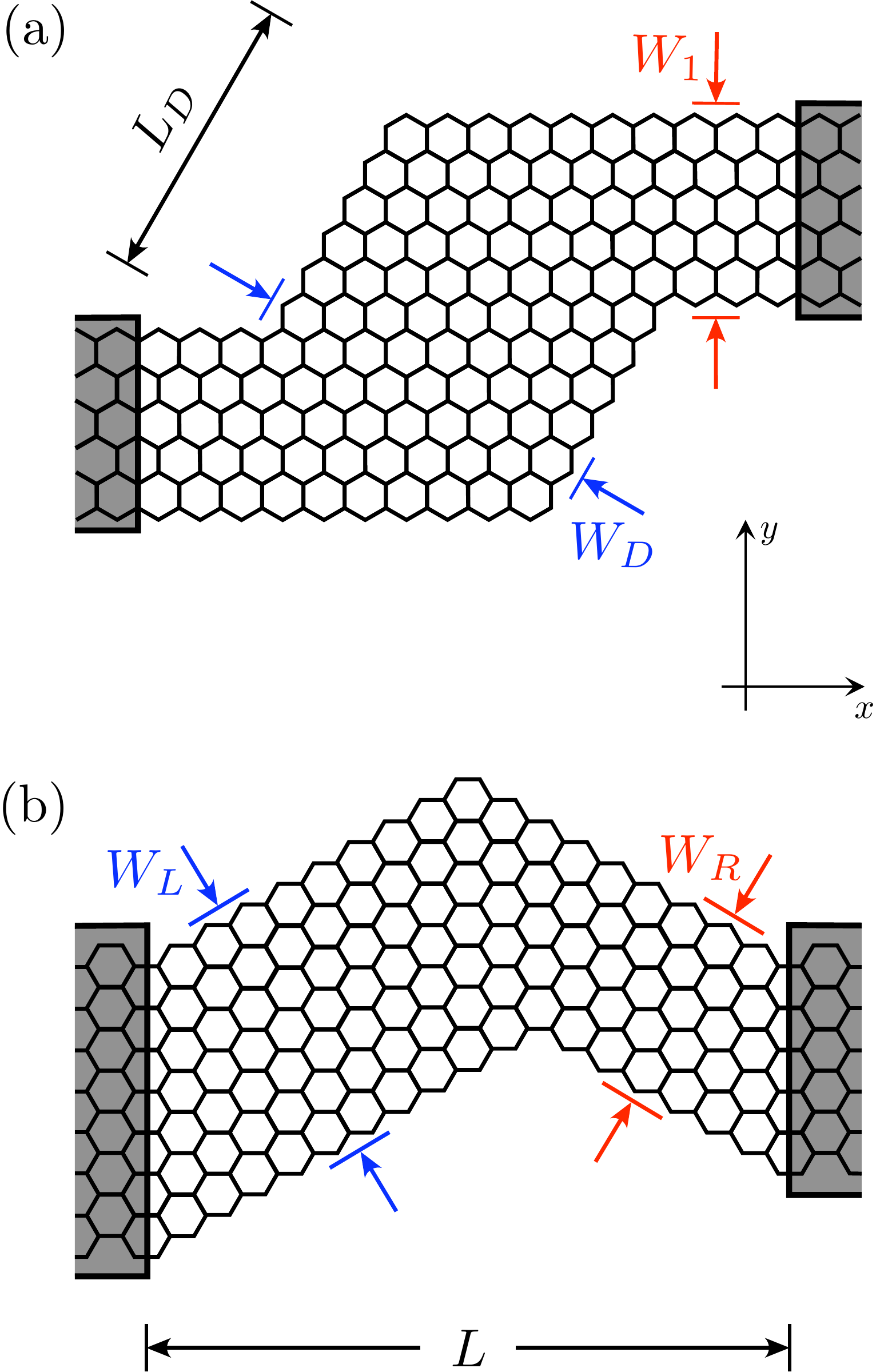}}
\caption{\label{kinks}
The systems studied numerically in the paper. (a) Quantum-dot device made of zigzag nanoribbon of the width $W_D$ and the length $L_D$, attached to similar ribbons of the width $W_1$ rotated by $60^\circ$. (b) The kink device formed by two zigzag ribbons of widths $W_L$ and $W_R$. Each of the systems is connected to heavily-doped graphene leads (shadow areas) displaced by the distance $L$ along the $x$-axis. }
\end{figure}

The device \cite{Wan07}, however, still needs sections of insulating ribbon with armchair edges, attached serially to both sides of the central section with zigzag edges (operating as SQD) to suppress the outward current flow. In this paper, we demonstrate numerically that a similar $Z$-shaped constriction, in which each of the three sections has zigzag edges (see Fig.\ \ref{kinks}(a)) and carries a fully conducting channel, may operate as highly-effective SQD due to mismatched valley polarization between the neighboring sections (rotated by $60^\circ$). The operation of such a double-kink device is rationalized by referring to the conductance spectrum of a single-kink device (Fig.\ \ref{kinks}(b)) which suppress the current \cite{Ryc08}, and by the discussion of level quantization for a~finite section of the ribbon with zigzag edges \cite{Mal08}.

The paper is organized as follows: In Sec.\ \ref{tbamod} we briefly present the numerical method of the conductance calculation in a tight-binding model of graphene, and discuss its relation to the effective Dirac equation. Then, in Sec.\ \ref{skink}, we recall the idea of valley polarization in constrictions with zigzag edges leading to the current suppression in the kink device, and calculate numerically the conductance of kink devices with various geometries. Finally, in Sec.\ \ref{dkink}, two kinks are attached serially and the resonance transmission via quantum-dot levels is discussed.

\section{Tight binding model for electron transport in graphene}
\label{tbamod}

We start from the nearest-neighbor tight binding model taking into account the $2p_z$ orbitals of carbon atoms \cite{Cas09}, with the Hamiltonian
\begin{equation}
\label{hamtba}
  H=\sum_{i,j}\tau_{ij}|i\rangle\langle j|+\sum_{i}V_{i}|i\rangle\langle i|.
\end{equation}
The hopping matrix element $\tau_{ij}=-\tau$ if the orbitals $|i\rangle$ and $|j\rangle$ are nearest neighbors on the honeycomb lattice (with $\tau=2.7\,$eV), otherwise $\tau_{ij}=0$. The electrostatic potential energy $V_{i}\equiv{V}(x_{i})$ varies only along the $x$-axis in the coordinate system of Fig.\ \ref{kinks}. It is chosen as $V_j=V_{\infty}$ in the leads marked by shadow areas ($x<0$ or $x>L$) or $V_j=0$ in the device area ($0<x<L$). For a given Fermi energy $E$, the chemical potential $\mu_j\equiv{E}-V(x_j)$ is equal to $\mu_\infty=E-V_\infty$ or to $\mu_0=E$, respectively. Throughout the paper, we analyze the system conductance as a function of $\mu_0$ at $V_\infty$ fixed such that $\mu_\infty$ corresponds to large number of propagating modes (the heavily-doped leads limit).

In the limit of zero bias voltage, phase-coherent transport properties of a noninteracting system such as described by the Hamiltonian (\ref{hamtba}) are encoded in the scattering matrix \cite{Dat97}
\begin{equation}
\label{smat}
S=\left(\begin{array}{cc}
  r & t' \\
  t & r' \\
\end{array}
\right),
\end{equation}
which contains the transmission $t$ ($t'$) and reflection $r$ ($r'$) amplitudes for charge carries incident from the left (right) lead, respectively. The conductance is determined by the Landauer-B\"{u}ttiker formula
\begin{equation}
\label{landa}
G=G_0\mbox{Tr}\,tt^\dagger=\frac{2e^2}{h}\sum_nT_n,
\end{equation}
where $G_0=2e^2/h$ is the conductance quantum, and $T_n$ is the transmission probability for the $n$-th normal mode. Apart from the simplest cases, when different transverse modes are not mixed by the transport \cite{Two06,Bre06} and the analytical solutions are available, one need to calculate the transmission matrix numerically. This can be achieved in two steps: First, we find the propagating modes in the leads $\Psi=\left[\psi_{n,+}^{(i)},\psi_{n,-}^{(i)}\right]$, where $\psi_{n,\pm}^{(i)}$ denotes the $n$-th incoming/outgoing mode in the $i$-th lead, with their self-energies $\xi_{n,\pm}$. Then, the scattering matrix (\ref{smat}) is obtained from the Lee-Fisher relation
\begin{equation}
\label{leef}
  S=-1-i\sqrt{v}\Psi^\dagger\mathcal{P}^\dagger\frac{1}{H-E+\Sigma}\mathcal{P}\Psi\sqrt{v}. 
\end{equation}
The self-energy term $\Sigma=\mathcal{P}\Psi\mbox{diag}(\xi_{n,+})\Psi^\dagger\mathcal{P}^\dagger$ represents the leads, with $\mathcal{P}$ the coupling matrix of the leads to the contact region. The matrix $v=\mbox{diag}(-2\mbox{Im}\xi_{n,+})$ contains normalization factors proportional to the propagating velocities for the modes. The direct matrix inversion in Eq.\ (\ref{leef}) is usually avoided by finding the $S$ matrix via the recursive Green's function algorithm, also available in a~version for a multi-terminal geometry \cite{Wim08}. For the two-terminal geometries considered here, and for the number of lattice sites $<10^5$, the direct matrix inversion with standard numerical routines is also very effective. According to Eq.\ (\ref{leef}), the transmission matrix $t$ depends on the properties of both constituents of the device under consideration: the leads and sample area. For the case of a weakly-doped graphene samples however, it was shown \cite{Sch07} that the transport of Dirac fermions is insensitive to the details of the leads, provided they carry a~sufficiently large number of propagating modes.

The method of the conductance calculation in a tight binding model of graphene, described briefly above, represents an adaptation---to the honeycomb lattice---of the method developed by Ando for a square lattice \cite{And91}. Albeit technically similar, from the fundamental point of view the two methods represents quite different approaches to mesoscopic physics. Ando's approach starts from a discretization of the Schr\"{o}dinger equation in order to solve the scattering problem for the geometries which are not tractable in a continuous limit. Here, we deal directly with the microscopic model of graphene, allowing one to extent the discussion on the situation when its effective model for low-energy excitations, given by the massless Dirac equation \cite{Cas09}, breaks down due to scattering the carriers between valleys \cite{Akh08b}. A finite-difference method for solving the Dirac equation on a square lattice was also recently developed \cite{Two08}.

\begin{figure}[!t]
\centerline{\includegraphics[width=0.7\linewidth]{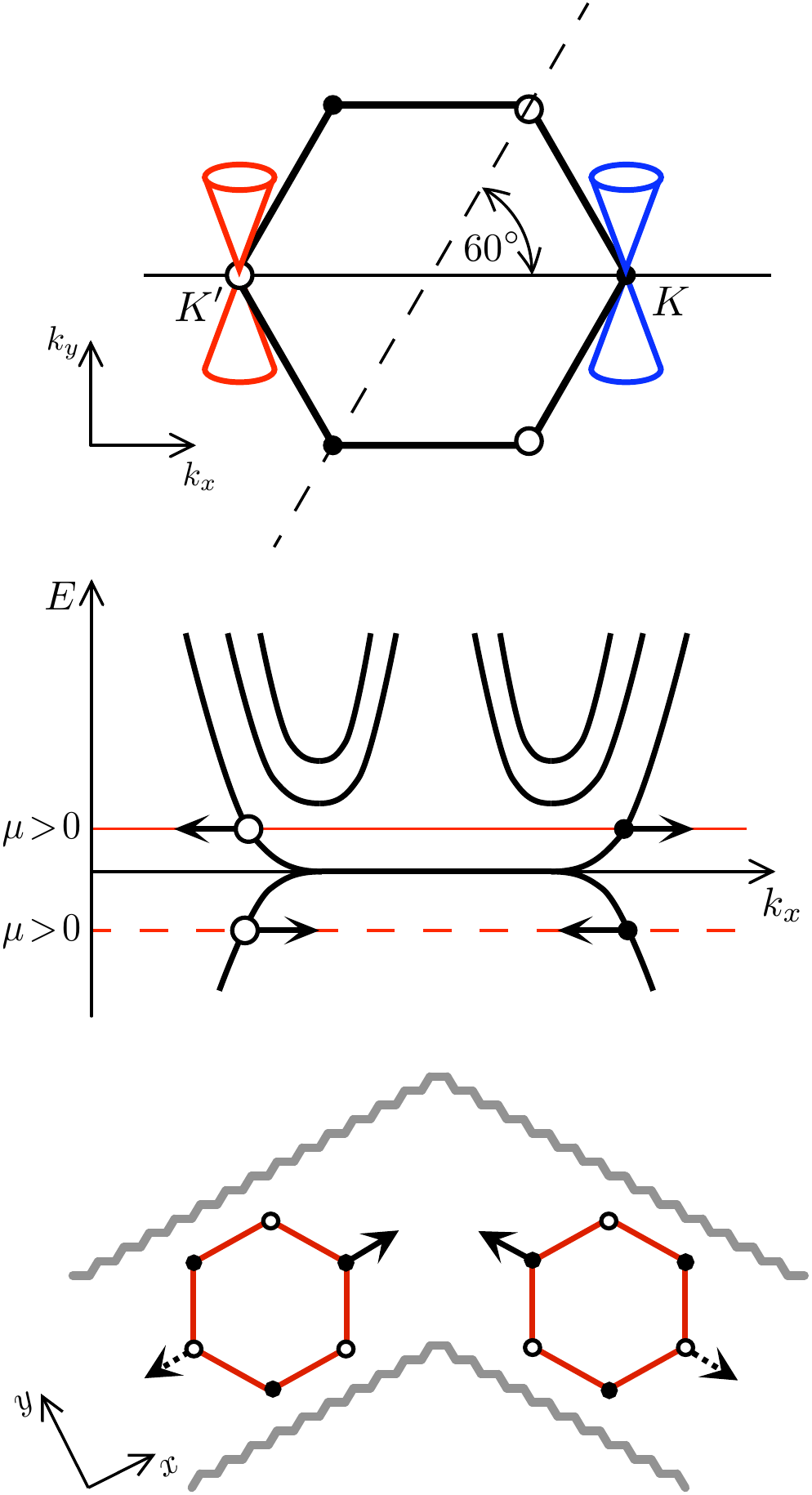}}
\caption{\label{valb}
Schematic illustration of the mechanism of current suppression in the kink device. Top panel: Hexagonal first Brillouin zone of graphene. The valleys centered at two inequivalent Dirac points, labeled $K$ and $K'$, are shown schematically. Middle panel: Dispersion relation for the ribbon with zigzag edges. For the lowest mode, there is one-to-one correspondence between the direction of propagation (indicated by arrows) and the valley isospin. The polarity switches $(i)$ when changes the sign of chemical potential $\mu$, or $(ii)$ when rotates the constriction axis by $60^\circ$. Bottom panel: First Brillouin zone at the two sides of the kink device. Solid (dashed) arrows indicate the direction of propagation for $K$ ($K'$) valley. As corresponding points in the Brillouin zone propagate in opposite directions at the two sides of the kink, electron cannot pass the device unless additional (microscopic) intervalley scattering processes are present. }
\end{figure}

\section{The kink device with zigzag edges}
\label{skink}
\subsection{Valley polarization and the electron transport}
To understand the physics ruling the conductance of devices shown in Fig.\ \ref{kinks}, one needs first to recall basic facts concerning electron transport through the constriction with parallel zigzag edges: a building block of each device considered. Such a constriction, also known as the \emph{valley filter} \cite{Ryc07}, was shown to produce, upon ballistic injection of current, the nonequilibrium valley polarization in a sheet of graphene attached. Motivated by the related analytical result for generic boundary conditions of graphene flakes \cite{Akh08a} we have shown numerically \cite{Ryc08}, that valley polarization is also produced by constrictions with other edges, apart from the perfect armchair edges. These observations are rationalized as follows: For a constriction long enough, the transport become one-dimensional, and crystallographic orientation of edges determines the direction of propagation in the first Brillouin zone (see Fig.\ \ref{valb}). As the lowest propagating mode in such a nanoribbon lack the twofold degeneracy of higher modes (and may be arbitrary close to one of the Dirac points $K$ or $K'$) the crystallographic orientation also determines the valley polarization of current passing the constriction (which is opposite for the conductance then for the valence band).

Now, the two quite different devices consisting of two valley filters (of the opposite polarity) in series, can be constructed. By simple applying a~step-like potential profile to the straight ribbon with zigzag edges, one obtains the electrostatically-controlled \emph{valley valve} \cite{Akh08b}. Alternatively, one can build the kink device \cite{Ryc08,Iye08}. In both cases, the current suppression due to mismatched valley polarization produced by two constituents of the device, is expected. However, a bit more detailed discussion of the current-suppression mechanism on the example of the kink device (see Fig.\ \ref{valb}, bottom panel) unveils its striking feature: Strictly speaking, a one-to-one correspondence between the direction of propagation and the valley index, appearing in a nanoribbon, causes the electron can neither be transmitted nor reflected by the interface between two filters of opposite polarity! The same observation applies to the valley valve, for which the theoretical analysis of microscopic tight-binding equations \cite{Akh08b} shows, that the intervalley scattering processes lead to sinusoidal conductance oscillations around the mean value $\overline{G}=0.5\,G_0$ when rotating the interface line with respect to the ribbon edge. The valve conductance also depends on its width, and is significantly different for the cases when zigzag and anti-zigzag ribbons are used, illustrating the microscopic nature of an electron transport. For the kink device, an analytical solution is missing, and the existing numerical results are overviewed briefly below.

\begin{figure}[!t]
\centerline{\includegraphics[width=0.9\linewidth]{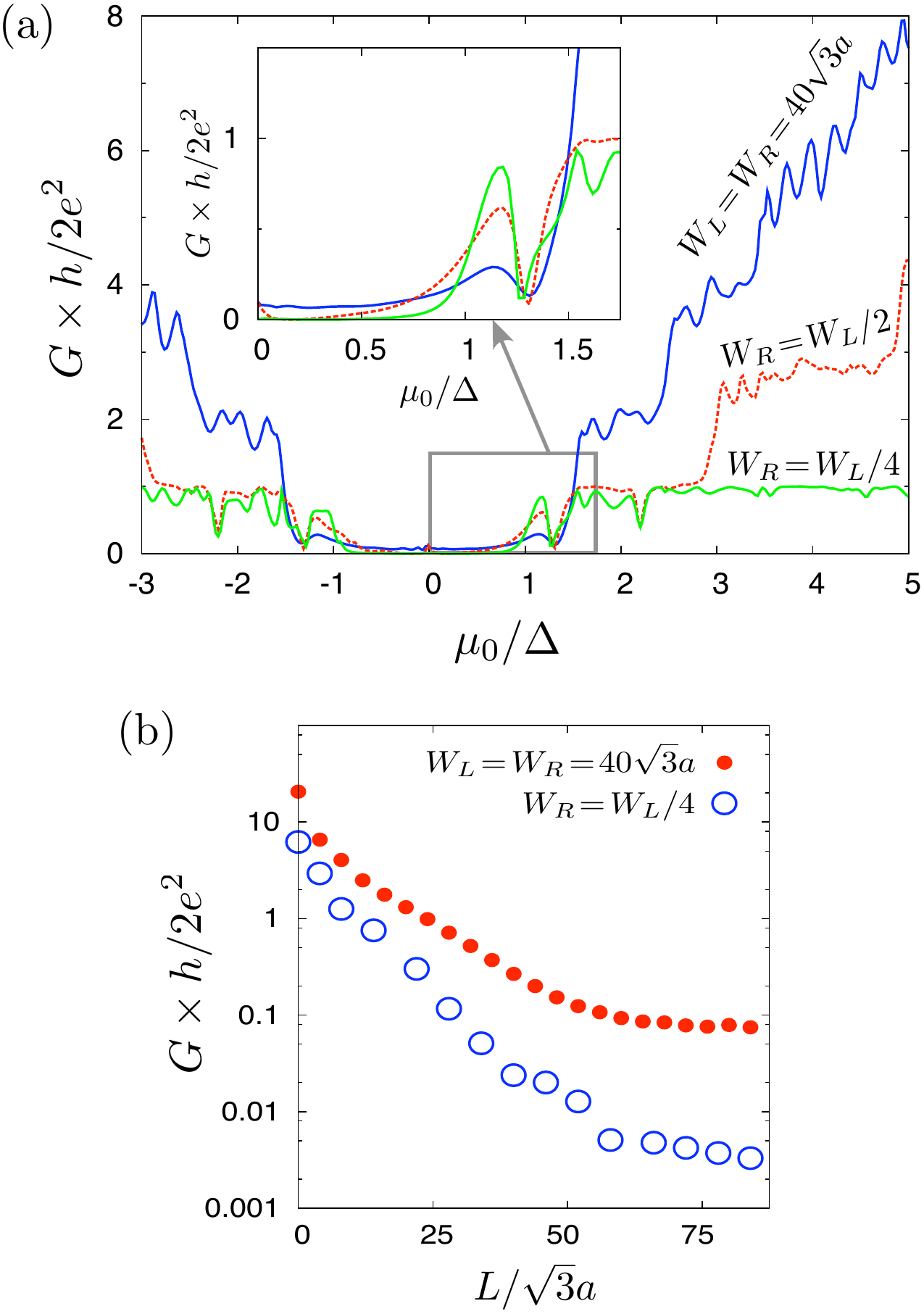}}
\caption{\label{gkin}
Conductance of the kink device as a~function of the chemical potential at fixed $L=100\sqrt{3}\,a$ (a) and as a function of the constriction length at fixed $\mu_0=\Delta/2$ (b). Inset in panel (a) shows a magnified region between the Dirac point and the first conductance step. In both panels, the width of left ribbon forming the kink is fixed at $W_L=40\sqrt{3}\,a$, whereas the width of right ribbon is varied between the curves. }
\end{figure}

\subsection{Conductance of the kink device}
The scattering problem for devices similar to that shown in Fig.\ \ref{kinks}(b) have been studied independently by several authors. The perfectly \emph{symmetric} kink, formed by two semi-infinite nanoribbons with zigzag edges each of which is carrying a~single propagating mode, shows irregular conductance oscillations with varying chemical potential \cite{Iye08}. The oscillations cover the full range of $0\leqslant{G}\leqslant{G_0}$, and the upper limit is approached when the resonances with quasi-bond states, localized at the kink symmetry axis, occur. The system of tight-binding equations describing the symmetric kink, however, was found to be numerically stable only for ribbons of a moderate width $W_L=W_R\lesssim{10}\sqrt{3}\,a$ (in units of the lattice spacing $a$). In the case of an \emph{asymmetric} kink with $W_L/W_R=2$ attached to heavily-doped graphene leads \cite{Ryc08} we have $G\ll{G_0}$ at the first conductance step, and the resonances do not appear. On the other hand, the kink-like system formed by joining the two ribbons rotated by $120^\circ$ shows almost a~perfect transmission, as the mismatched-valley polarization does not appear in such a~case \cite{Are07}.

Here, we consider a slightly different geometry then studied in Refs.\ \cite{Ryc08,Iye08}. Namely, the kink device is attached to heavily-doped graphene leads with armchair edges. This guarantees the reflection symmetry of the system when $W_L=W_R$, and eliminates the problem with numerical stability occurring for wider ribbons in the setup of Ref.\ \cite{Iye08}. We have fixed the left ribbon width at $W_L=40\sqrt{3}\,a$, corresponding to the subband splitting $\Delta\equiv\frac{1}{2}\sqrt{3}\,\pi\tau a/W_L=\pi\tau/80$ and $\nu_L=32$ propagating modes for $\mu_\infty=V_\infty=\tau/2$. The right ribbon width is varied as ${W_R}/(\sqrt{3}\,a)=10$, $20$, and $40$. 

The operation of the kink device is demonstrated in Fig.~\ref{gkin}. First, we took a~sample area of the length $L=100\sqrt{3}\,a$ and plot, in Fig.~\ref{gkin}(a), the conductance as a function of the chemical potential. All three devices with different $W_R$ show some conductance suppression at the first plateau $|\mu_0|\lesssim\Delta$. The conductance of the symmetric kink ($W_R=W_L$) is still of the order of $G\sim{0.1}\,G_0$, but the asymmetric kinks shows much stronger suppression at low energies (see the inset). For an additional illustration we present, in Fig.\ \ref{gkin}(b), the conductance as a~function of $L$ at fixed $\mu_0=\Delta/2$. For $L=0$, we have $G/G_0\approx\mbox{min}(\nu_L,\nu_R)=\nu_R$, with the number of propagating modes in the right arm $\nu_R=32$ or $8$ (for $W_R=W_L$ or $W_L/4$, respectively). For larger $L$, $G$ first decrease exponentially, in order to saturate for $L\gtrsim\sqrt{3}{W_L}/2$ (the length above which the role of evanescent modes becomes negligible) in the symmetric case. For $W_R=W_L/4$, $G$ continues to decay and approaches the value $\sim{10}^{-3}G_0$ for longest examined systems, confirming the claim of Ref.\ \cite{Ryc08}, that the asymmetric kink device in graphene blocks the current very effectively at low dopings. These findings also coincides with earlier results for Aharonov-Bohm quantum rings \cite{Rec07,Ryc09} for which intervalley scattering rate was shown to be related to the presence of an asymmetry or edge irregularities. 

Away from the first plateau, each device shows conductance quantization that is essentially governed by the narrower arm width $W_R$. In the symmetric case, however, we have the steps corresponding to \emph{even} multiplicities of $G_0$, whereas in the asymmetric case \emph{odd} multiplicities appear. This can be summarize in an approximating formula for the kink conductance 
\begin{equation}
\label{gpprx}
G/G_0\approx\mbox{min}(\nu_L,\nu_R)-\delta_{\nu_L,\nu_R},
\end{equation}
where the Kronecker delta  $\delta_{\nu_L,\nu_R}$ accounts for mismatched valley polarization of the lowest propagating modes in two kink arms, which affect the total conductance also at higher dopings.

\begin{figure}[!t]
\centerline{\includegraphics[width=0.9\linewidth]{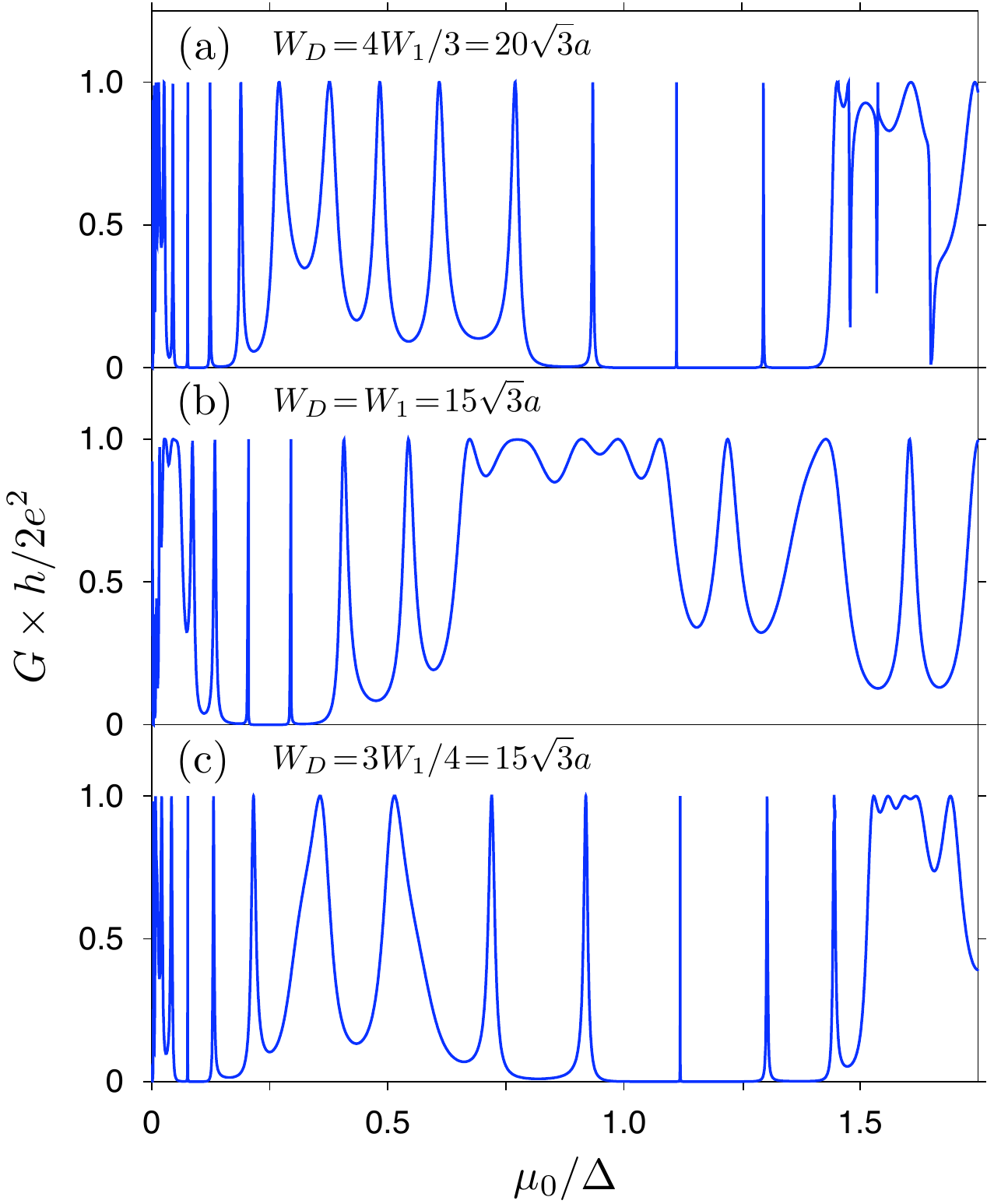}}
\caption{\label{conki}
Conductance of the double-kink device for different arrangements of its ribbon-like sections. (a) Narrow-wide-narrow, (b) uniformly-wide, and (c) wide-narrow-wide setup. }
\end{figure}

\begin{figure}[!t]
\centerline{\includegraphics[width=0.9\linewidth]{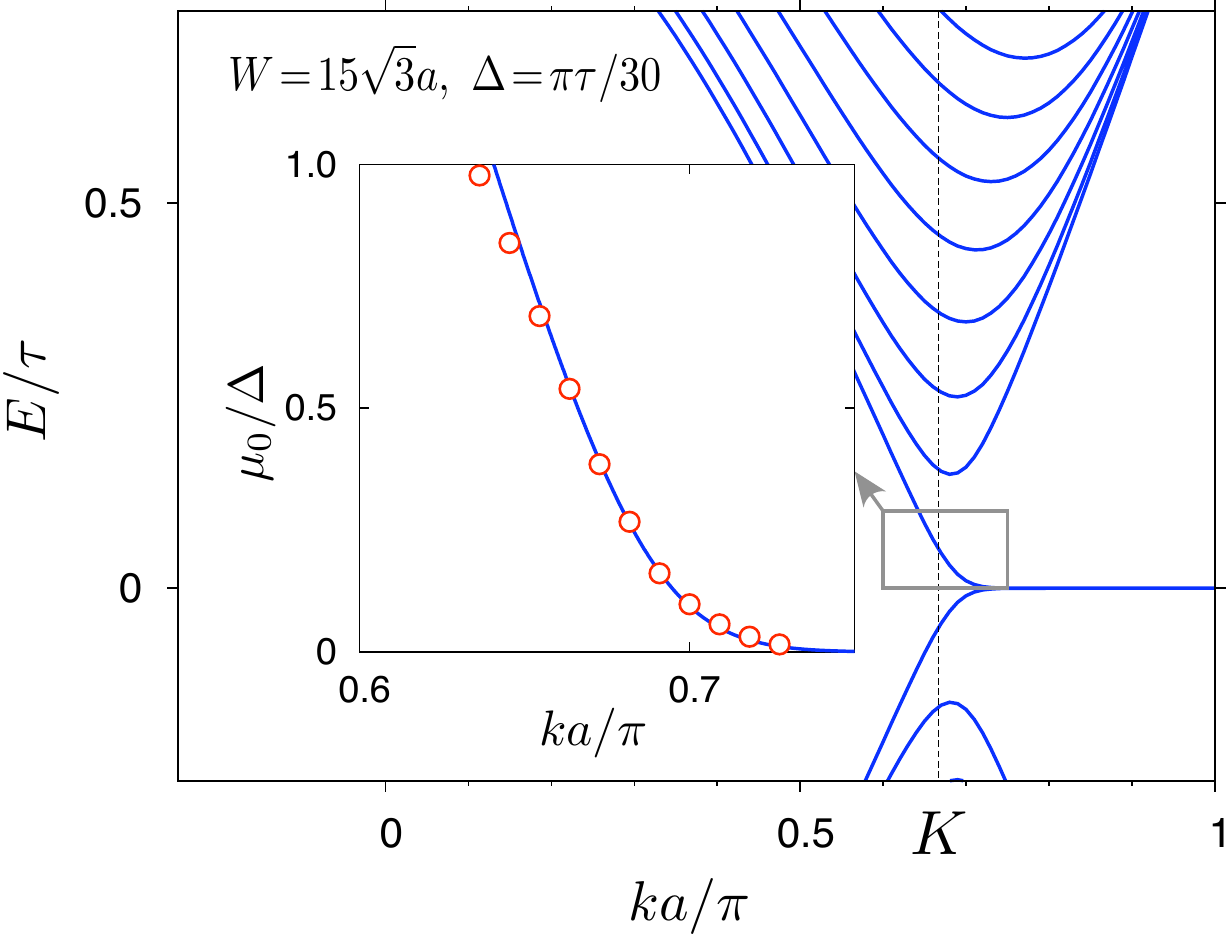}}
\caption{\label{peaks}
Quantum-dot energy levels of the double-kink device. Lines: dispersion relation of a nanoribbon with zigzag edges. Datapoints in the inset: conductance maxima of a double kink with $W_D=15\sqrt{3}a$, $W_1=4W_D/3$, and $L_D=110a$. } 
\end{figure}

\section{Bound states of a double kink}
\label{dkink}
The conductance spectra of $Z$-shaped nanoconstriction of Fig.\ \ref{kinks}(a) are presented in Fig.~\ref{conki}. We consider three different devices, all built by joining finite sections of two nanoribbons with zigzag edges: one of the $15\sqrt{3}\,a$ width, and the other of the $20\sqrt{3}\,a$ width. In the first case, we arrange the building blocks in a double-kink setup, such that the central section is wider then the peripheral sections ($W_D=20\sqrt{3}\,a$, $W_1=15\sqrt{3}\,a$). Next, we consider the setup of a uniform width ($W_D=W_1=15\sqrt{3}\,a$). Finally, we took the peripheral sections wider than the central one ($W_D=15\sqrt{3}\,a$, $W_1=20\sqrt{3}\,a$). For each setup, the total sample area length is fixed at $L=160\,a$. The results show, either for the narrow-wide-narrow or for the wide-narrow-wide setup (see Figs.\ \ref{conki}(a) and \ref{conki}(c), respectively), that $G\approx{0}$ at the first conductance plateau except from the narrow peaks, corresponding to resonances with the quantum-dot states of the central section. Only for a uniformly-wide device some finite intervals of $\mu_0$, for which $G\leqslant{G_0}$, appear in Fig.\ \ref{conki}(b). 

These findings are related to the operation of the kink device studied in previous section as follows: The asymmetric kink blocks the current at low dopings, so two identical devices of its kind in series function as an electrostatically-controlled quantum dot (with an electron trapped between the kinks) regardless we join together wider, or narrower of the kink arms. The two symmetric kinks in series may also trap an electron accidentally, when the eigenstate of the central section lies in an energy interval, for which the single-kink conductance is low enough.

In Fig.~\ref{peaks}, we plot the positions of the conductance maxima extracted from Fig.\ \ref{conki}(c), together with the dispersion relation for an infinitely long nanoribbon of $W=15\sqrt{3}\,a$ width. The $k$-coordinate of each maximum is related to its number in sequence $j$ as $k_j=\pi(j_\star-j)/L_D$, with $j_\star$ determined via $E(K)=\Delta/2$. (Notice that the first conductance maxima are blurred at finite machine precision due to the dispersionless character of the lowest subband). Strictly speaking, each bound state of the central ribbon section of $L_D$ length is given by a~unique quantum superposition of propagating waves from $K$ and $K'$ valleys \cite{Mal08}, the dispersion relation for one valley is, however, a~mirror reflection of that for the second valley. The datapoints shown in the inset of Fig.~\ref{peaks} follow the dispersion relation for the lowest subband of a nanoribbon with zigzag edges. Hence, for each peak in the conductance spectrum of a double-kink device, the corresponding quantum-dot energy level is found.

\section{Conclusions}
In conclusion, we have studied numerically the electron transport through $Z$-shaped nanoconstriction with zigzag edges, operating as a~quantum dot in graphene. The device conductance, analyzed as a~function of the chemical potential, exhibits a~series of narrow resonance peaks. Each of the resonances is linked up to the energy level of the central section, when separated from the other parts of the device. An electron-trapping mechanism is discussed by analyzing the transport through a basic device building block: the kink with zigzag edges. A~moderate current suppression is observed for the symmetric kink, whereas the asymmetric kink was found to reflect electrons almost perfectly. The role of a~mismatched valley polarization on both sides of the kink is stressed. 

Similar bound states were recently found in the simulation of transport through $S$-shaped nanoribbon \cite{Wur09b} with irregular edges. However, such a~system contains finite sections of a~nanoribbon with armchair edges, so the nature of the electron-trapping mechanism is not as clear as for the simpler system considered here.

\section*{Acknowledgment}
I thank to Anton Akhmerov, Patrik Recher, C.W.J.\ Beenakker, \.{I}nan\c{c} Adagideli, Klaus Richter, Michael Wimmer, and J\"{u}rgen Wurm for helpful discussions and correspondence. The work was mainly completed during the Alexander von Humboldt (AvH) fellowship in Regensburg. Partial supports from the Polish Science Fundation (FNP) and the Polish Ministry of Science (Grant No.\ N--N202--128736) are acknowledged.

\end{document}